\documentclass[10pt, a4paper, reqno]{article} %Touati
\usepackage{amsmath,amssymb,bm,yhmath}
\usepackage[all]{xy}
\usepackage{caption,graphicx}
\usepackage{geometry}
\begin{document}

\title {Remembering Yvonne Choquet-Bruhat}

\author{Thibault Damour (IHES)}

\maketitle
  
\begin{abstract} I describe  the impact of some  of the mathematical results  of Yvonne Choquet-Bruhat on gravitational physics,
as well as the evolution of my interactions with her over the years.
 \end{abstract}

 \section{First encounter}
 
My earliest memory of Yvonne Choquet-Bruhat (whom I will simply call Yvonne) is her smiling face when we first met on a train platform in Paris. Was it July 1975, on our way to the first Marcel Grossmann Meeting in Trieste? I don't know, but I do clearly remember her warm, welcoming smile toward a junior scientist just starting out in gravitational physics. That smile never faded throughout all the years I had the pleasure of interacting with her -- particularly between 2003 and 2018, when she was a regular visitor at IHES.

 \section{First interactions linked to science}
 
 The Comptes Rendus of the (French) Academy of Sciences  were
originally created to announce, briefly and rapidly, significant new results. In particular, they allowed junior
scientists to publish a short account of novel results under the aegis of a member of the Academy of Sciences
who served as a ``referee", vouching for the (a priori) validity of their results.
For instance, Yvonne's first announcement of her breakthrough result on  Einstein's field equations, at the early age of 26,
namely her ``Th\'{e}or\`{e}me d'existence pour les \'{e}quations de la gravitation einsteinienne dans le cas  non analytique", was published in early 1950,  a  couple of weeks after the presentation of her Note to the Academy, on February 6, 1950,  by Jacques Hadamard (one of the greatest living mathematicians at the time).
See \cite{YCB50} for an English translation, and republication, of  Yvonne's Note as a Golden Oldie, 
and \cite{Christodoulou:2022ijn} for a beautiful accompanying editorial comment .

I had become aware of this possibility of ``recording the date" (``prendre date", in the words of Lichnerowicz)
 in the fall of 1974  when Remo Ruffini introduced me to Andr\'e Lichnerowicz
in Princeton. [And this eventually led to the fast publication (December 1974) of my first published paper, a Comptes Rendus 
Note with Remo on the newly discovered Hulse-Taylor binary pulsar \cite{DamourRuffini}.] 
The reason for my mentioning this way of helping junior scientists to publish rapidly novel results (and also to bring their
work to the attention of senior experts in their field) is that, around 1980, when Nathalie Deruelle and I obtained
novel results on the two body problem in General Relativity we asked Yvonne to present to the Academy two
short Notes \cite{DD81a,DD81b} on the Lagrangian dynamics of two point masses at the second post-Newtonian approximation,
soon followed by another Note by me \cite{D82} also presented by Yvonne.
Nathalie and I were quite grateful to Yvonne for this triple opportunity of expeditiously announcing our new results
at a moment where there was an international, competitive effort to clarify the so-called ``quadrupole 
formula controversy", i.e. the issue of understanding the back reaction of gravitational radiation on the
dynamics of binary systems, such as the Hulse-Taylor binary pulsar.

 \section{Deeper scientific interactions with Yvonne}
 
 Though, as mentioned above, I had met Yvonne early in my career, and I continued meeting her,
 in an increasingly friendly way, in many subsequent scientific meetings (Marcel Grossmann meetings, 
 Journ\'ees Relativistes \cite{JR}, GRG meetings, etc.), I must confess that, for many years,
 I had no real scientific interactions with her. Seen from the perspective of a junior theoretical physicist,
 interested in exploring the observable consequences of Einstein's gravitation theory 
 in our real Universe, most of her work appeared to me as being too mathematically-oriented to 
 attract more than a rather  far-away contemplation. It took me years to understand that I was wrong,
 and that Yvonne's works contained many results of direct importance for gravitational physics.
 
 Let me  mention just a few of the physics-relevant results of Yvonne (besides her pioneering
 non-analytic existence theorem \cite{YCB50,YCB52} whose physics importance I also only slowly appreciated,
 and her many results on the constraints, which, I assume, are discussed in other contributions to this homage):

 \begin{itemize}
 
\item  First, her pioneering work \cite{YCB56} on the $3+1$ decomposition of Einstein's equations, usually attributed only
 to Arnowitt-Deser-Misner \cite{ADM}. As we know, the $3+1$ decomposition of Einsteinian gravity has become
 one of the key elements of modern Numerical Relativity. More about this below.
 
\item  Second, her work on strong high-frequency gravitational waves \cite{YCB69}. It took me time to grasp what
 made this work significantly superior to the famous, slightly earlier, work of Rich Isaacson \cite{Isaacson:1968zza}.
 Besides a mathematically clear way of defining (by the two-timing technique) and justifying (by a boundedness condition) the averaging leading to the appearance of the effective stress-energy tensor of gravitational waves,
 the most interesting result of \cite{YCB69} is the realization that the nonlinear structure of Einstein's equations
 ensures the lack of steepening of strong gravitational waves during their propagation. In other words, nonlinear
 gravitational waves satisfy the exceptionality condition of Lax and Boillat (see, e.g., \cite{Boillat:1969slt}
 and references therein). 
 This exceptional property of nonlinear gravitational waves is directly related to the weak form of the
 Christodoulou-Klainerman ``null condition" satisfied by Einstein's equations (see \cite{Lindblad03} and references
 therein).  For recent progress in the construction of exact solutions containing strong high-frequency gravitational
 waves see \cite{Touati:2024xpw} and references therein.
 
\item  Then I would like to mention Yvonne's work on the positive mass issue. Most importantly, her 1976 work
with Jerrold Marsden \cite{YCB76} gave the first rigourous proof of the positivity of energy for vacuum 
space times near flat space using a critical point analysis in infinite dimensions.  See below for a later
contribution of Yvonne, and \cite{Huang25} for a recent survey of positive(-mass)-energy theorems.
 
\item Of particular importance is the work of Yvonne on hyperbolic formulations of 3+1 versions
of the evolution part of Einstein's equations.  In particular, Yvonne and Tommaso Ruggeri obtained
(by combining the evolution equations with the constraints) a 3+1 hyperbolic system with zero shift $\beta$.
This was then generalized by Yvonne (in collaboration with  Jimmy York and Arlen Anderson) to the construction
of hyperbolic 3+1 systems with non-zero shift. These results were part of an international effort to construct
3+1 evolution systems having sufficient stability to be solved on a computer. Let me note in particular that the
crucial time-hyperbolicity condition used by Yvonne and Tommaso Ruggeri, which yields, in 3+1 variables,
the following evolution equation for the lapse $\alpha$, $(\partial_t - {\mathcal L}_{\beta})= - \alpha^2 K$,
was later generalized   \cite{Anninos:1995am} in the so-called ``1+log" slicing condition, 
$(\partial_t - {\mathcal L}_{\beta})= - 2 \alpha K$ which has become an important ingredient of many
modern Numerical Relativity codes.
 
\item  Let me finally mention the important contributions of Yvonne to understanding the
causality properties of several of the extensions of Einstein's theory suggested by modern theoretical physics.
I have notably in mind here the works of Yvonne on Supergravity \cite{Choquet-Bruhat:1983xyr,Bao:1984bp,Choquet-Bruhat:1985xei} and on Gauss-Bonnet-type gravity \cite{Choquet-Bruhat:1988jdt}.

\end{itemize}
  
  \section{Discussions with Yvonne at IHES}
  
  From 2003 to 2018, Yvonne regularly visited  IHES (at least once per week), 
  and I had the pleasure to discuss many times
  with her, about either mathematics or physics.  This also allowed me to have many enlightening discussions 
  with several of the collaborators of Yvonne who visited IHES in those years, notably Vince Moncrief
  and Piotr Chrusciel. Let me first recall that I was always amazed by the clearness and accuracy of her answers
  to any of my math questions. She would either immediately answer a math question with a precise answer,
  or give me the name of the best person to contact.
  
  During her stay at IHES, Yvonne wrote some twenty-six research papers, two highest-level scientific books,
  \cite{Choquet-Bruhat:2009xil} and \cite{Choquet-Bruhat:2014okh}, and her autobiogaphy \cite{YCBbio}.
  I had uncountable discussions with her about the content of all her books. 
She kindly asked me to write a chapter for her monumental, testamentary book \cite{Choquet-Bruhat:2009xil}
on the Belinsky-Khalatnikov-Lifshitz conjecture concerning the  behavior of generic solutions of Einstein's equations
near a spacelike singularity. I was happy to do so, and I tried to formulate, in an as mathematically precise
way as I could, the conjecture suggested in the pioneering work of
Belinsky, Khalatnikov and Lifshitz \cite{Belinsky:1970ew}, and refined in
 the many physics works triggered by it. Let me note in this respect the not so well-known fact
that the main results of \cite{Belinsky:1970ew}, 
and notably the coupled second-order nonlinear differential equations
for the three local scale factors $a(\tau), b(\tau), c(\tau)$, were first publically presented by
Isaak Khalatnikov in January 1968 at the Institut Henri Poincar\'e in Paris \cite{Khalat}. The audience comprised in particular,
John Archibald Wheeler and Demetrios Christodoulou. When Isaak presented the $a-b-c$ ODEs, Wheeler
made the public remark that this looked like a mechanical system where the Universe was described by the
three coordinates $a, b, c$. One year later (14 April 1969), Charles Misner published his famous Lagrangian
description of the  complex dynamics of Bianchi IX universes (called by him ``Mixmaster Universe"), mentioning
at the end a private communication from J.A.W. in which he ``suggested that studies of singularities by 
Belinsky and Khalatnikov had also found alternating Kasner-like epochs but with very simple description
 in terms of a related parameter ($u$)".

Her second scientific book \cite{Choquet-Bruhat:2014okh} led also to many interesting discussions with Yvonne,
who wanted to be kept abreast of the points of contact between General Relativity and experimental or observational
facts. Let us recall in this respect that Yvonne (whose father, Georges Bruhat, was a physicist, well known in France for his work on optics
and his textbooks) considered herself as a ``failed" physicist who constantly aimed at understanding the real universe through its theoretical physics description, by using, and perfecting, mathematical tools.

During Yvonne's stay at IHES, two conferences were organized in her honor: one in March 2004 for her 80th birthday, and
one in January 2014 for her 90th birthday. Another special moment happened on February 11, 2016:  Yvonne and a group of scientists of IHES (including me) followed the live announcement
(from Washington, DC, USA) of the discovery of the first gravitational wave signal by the two LIGO interferometers.
After the end of her regular visits at IHES,  a one-day conference 
was organized at IHES  in December 2023 to celebrate her 100th birthday (unfortunately in her absence).

Let me finally mention that I had detailed technical discussions with Yvonne about one of her later research papers,
namely her  streamlined, complete proof (valid in arbitrary space dimension $n$, and using only spinors on some (oriented) spacelike section $\Sigma_n$)  of
the (strong) positive energy theorem ($ E \geq |{\bf P}|$) in General Relativity. She kindly offered me to co-sign her
paper. But I felt I had not significantly contributed to her proof, and I declined this honor, and this token of friendship.

 \section{Epilogue}
 
 To end this tribute to the memory of Yvonne, let me mention one of my last discussions (by phone) with her.
 One day in the spring, or early summer, of 2018 (at a time when she had essentially stopped visiting
 regularly IHES), Yvonne called me on my cell phone (I remember that I was
 walking in the countryside) to tell me the good news that the  Italian Society of General Relativity and Gravitation 
 (SIGRAV) had contacted her to inform her that she had been awarded  the 2018 Amaldi Medal,  to be received in
 person at the next SIGRAV National conference, due to take place 
 in September  in Cagliari (Sardegna, Italy).
 She was very happy to receive this distinction because she always had had close links with Italy and very
 friendly relations with many Italian scientists.
 However, she told me that she did not feel she had the energy to go there in person (she was 95 !). 
 I offered to collect the medal on her behalf, and to make a small presentation of her life work. 
 So it happened. It gave me the opportunity to study in
 detail some of her  most important contributions to gravitational physics, which I greatly enjoyed.
 [However, as I had to travel after the SIGRAV meeting to a summer school
 in Ravello, I had the  slight practical problem to be cautious in carrying all over Italy this very  valuable Amaldi medal,
 made of solid gold!]
 
%\section*{Acknowledgements}
%I  express my gratitude to Nathalie Deruelle for her helpful remarks.

 \end{document}